\documentclass[reprint,superscriptaddress,prb,aps]{revtex4-2}

\usepackage{mathtools}
\usepackage{amsmath}
\usepackage{amsthm}
\usepackage{amssymb}
\usepackage{graphicx}
\usepackage{bm}
\usepackage[colorlinks,linkcolor=blue, urlcolor=blue,citecolor=blue,bookmarks=false]{hyperref}

\begin{document}

\title{Doublon-like excitations and their phononic coupling \protect\\ in a Mott charge-density-wave system}

\author{C. J. Butler}
\email{christopher.butler@riken.jp}
\affiliation{RIKEN Center for Emergent Matter Science, 2-1 Hirosawa, Wako, Saitama 351-0198, Japan}

\author{M. Yoshida}
\affiliation{RIKEN Center for Emergent Matter Science, 2-1 Hirosawa, Wako, Saitama 351-0198, Japan}

\author{T. Hanaguri}
\email{hanaguri@riken.jp}
\affiliation{RIKEN Center for Emergent Matter Science, 2-1 Hirosawa, Wako, Saitama 351-0198, Japan}

\author{Y. Iwasa}
\affiliation{RIKEN Center for Emergent Matter Science, 2-1 Hirosawa, Wako, Saitama 351-0198, Japan}
\affiliation{Quantum-Phase Electronics Center and Department of Applied Physics, The University of Tokyo, 7-3-1 Hongo, Bunkyo-ku, Tokyo 113-8656, Japan}

\begin{abstract}
Electron-phonon-driven charge density waves can in some circumstances allow electronic correlations to become predominant, driving a system into a Mott insulating state. New insights into both the Mott state and preceding charge density wave may result from observations of the coupled dynamics of their underlying degrees of freedom. Here, tunneling injection of single electrons into the upper Hubbard band of the Mott charge-density-wave material 1\textit{T}-TaS$_{2}$ reveals extraordinarily narrow electronic excitations which couple to amplitude mode phonons associated with the charge density wave's periodic lattice distortion. This gives a vivid microscopic view of the interplay between excitations of the Mott state and the lattice dynamics of its charge density wave precursor.
\end{abstract}

\maketitle

\section{Introduction}
Coupling between electronic and lattice degrees of freedom underpins many intriguing and useful phenomena in condensed matter systems, such as charge density waves (CDWs) and the pair-binding mechanism in conventional superconductivity \cite{Wilson1975,Johannes2008,Bardeen1957,Eliashberg1960}. The layered transition metal dichalcogenide 1\textit{T}-TaS$_{2}$ is a material whose rich electronic phase diagram and ground state electronic properties are shaped principally by complex electron-phonon (\textit{e-ph}) interactions. At low temperature, \textit{e-ph} interactions drive a commensurate CDW described by a $\sqrt{13}\times\sqrt{13}$ \textit{R}13.9$^\circ$ superlattice. Figure 1(a) shows a typical scanning tunneling microscopy (STM) image acquired in this phase, at a temperature of 1.5~K. This CDW can equivalently be described as a lattice of polaronic Star-of-David (SD) clusters each encompassing 13 Ta ions \cite{Wilson1975} and formed from the symmetrical contraction of twelve of the Ta ions towards a central one, as depicted in Fig. 1(b).

The most widely adopted understanding of the ground state in 1\textit{T}-TaS$_{2}$ is that the CDW reduces the electronic bandwidth below the threshold for a Mott transition into an insulating state \cite{Fazekas1979,Fazekas1980}. Nevertheless much is yet to be clarified about the driving mechanisms behind both the CDW itself and the Mott state, and about the detailed interplay between \textit{e-ph} interactions and electronic correlations. Time-resolved photo-excitation measurements have provided fresh insights by investigating the system's dynamics, from excitation of the CDW amplitude mode \cite{Perfetti2008} and doublons (characteristic excitations of the upper Hubbard band (UHB) of a Mott insulator) in the weak-perturbation regime \cite{Mann2016,Ligges2018}, to timing the melting of both the Mott state and the CDW towards metallic and other phases under high-intensity light \cite{Petersen2011, Hellmann2012, Zhang2019, Avigo2019}.

Here we report on a spectroscopic signature of the dynamic coupling between electronic and lattice degrees of freedom in 1\textit{T}-TaS$_{2}$, with the electronic perturbation provided not by photo-excitation, but instead by local tunneling injection of single electrons using scanning tunneling microscopy (STM). Specifically, injection of electrons into unusual and previously unobserved narrow states found at the UHB onset likely excite amplitude mode phonons associated with the CDW lattice distortion. The narrow states themselves may evidence either long-lived coherent quasiparticles existing at the UHB edge, or possible polaronic bound states formed upon injection.
These microscopic observations complement the ensemble-averaged ultra-fast optical and time-resolved photo-emission spectroscopy measurements and also provide a reference point for theories modeling exotic excitations of Mott-Hubbard and related systems.

\begin{figure*}
\centering
\includegraphics[scale=1]{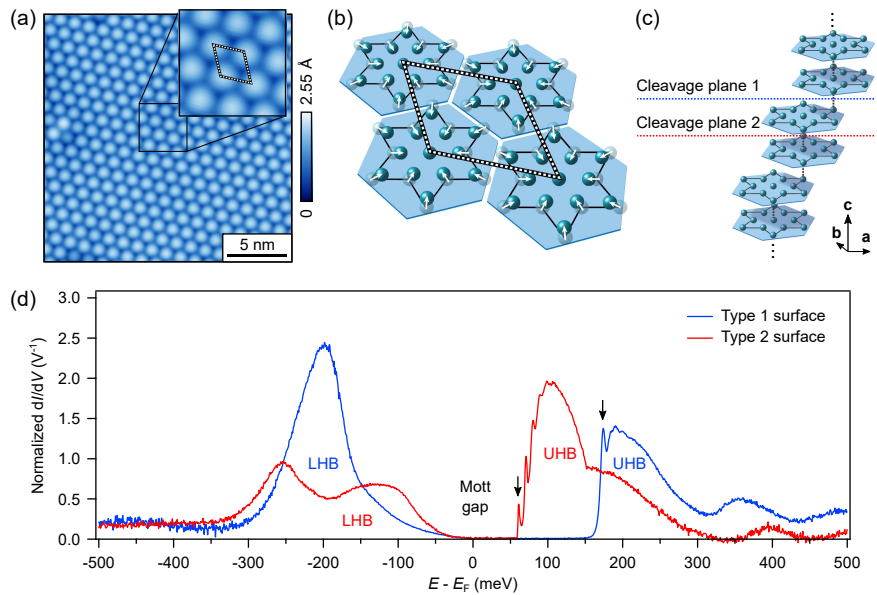}
\caption{\label{fig:1} Spectroscopically distinct surface terminations of the three-dimensional charge order in 1\textit{T}-TaS$_{2}$. (a) A typical constant-current topograph ($V$ = 250~mV, $I$ = 250~pA), acquired at a Type 1 surface. (b) A depiction of the periodic lattice distortion forming SD clusters with 13 Ta atoms each (sulfur ions are neglected). A surface-projected supercell of the CDW distortion is shown with a black \& white dashed line, corresponding to the cell shown in the inset of (a). (c) The alternating inter-layer stacking of the SD clusters, showing the two inequivalent cleavage planes 1 \& 2. (d) High-resolution $\frac{\textrm{d}I}{\textrm{d}V}(V)$ curves acquired at two types of surface. Recent STM investigations allow the identification of each spectrum with the Type 1 \& 2 surfaces formed by cleavage through the planes 1 \& 2 in Fig. 1(c) \cite{Butler2020}. The newly observed peaks at each surface are indicated with arrows. For ease of comparison the curves are normalized by dividing by the current value at +500~mV for each.}
\end{figure*}

\section{Results}
High-energy-resolution tunneling conductance ($\frac{\textrm{d}I}{\textrm{d}V}$) spectra were acquired at two distinct surface terminations of the three-dimensional CDW pattern. The observation of two surfaces is relevant as follows: The status of 1\textit{T}-TaS$_{2}$ as a Mott insulator was recently challenged \cite{Ritschel2018,Lee2019} with the suggestion that staggered inter-layer stacking \cite{Tanda1984,Naito1984,Naito1986} could result in a simple band insulator through Peierls-like inter-layer dimerization.  There is now mounting evidence from state-of-the-art X-ray diffraction \cite{Stahl2019}, high- and low-energy electron diffraction \cite{LeGuyader2017,vonWitte2019} and STM experiments \cite{Butler2020} that such an alternating stacking pattern is indeed realized in the low-temperature phase, and the band insulating picture has attracted some support \cite{Wang2020}. Nevertheless, we recognize that Peierls-like dimerization and an electron-correlation driven insulating state are not mutually exclusive \cite{Fuhrmann2006}, and behaviors consistent with a correlation-driven insulator continue to be reported \cite{Ligges2018,Bu2019,Butler2020}.  As depicted in Fig. 1(c), the stacking pattern features two inequivalent cleavage planes, labeled as planes 1 \& 2, forming dimerized or un-dimerized surfaces which yield distinct tunneling spectra of Type 1 or 2 [blue and red curves in Fig. 1(d)] respectively. As we suggested recently, if the inter-layer stacking pattern is truncated at a surface (namely, of Type 2) leaving an un-dimerized layer, the persistence of a gap in absence of dimerization suggests electronic correlations as the determining mechanism \cite{Butler2020}. Hence, in this work we adopt the viewpoint that the major spectroscopic features described below can be interpreted in the Mott-Hubbard picture. 

Particularly noteworthy in these spectra are the exceptionally narrow peaks appearing at the onset of unoccupied states at each surface, marked with black arrows in Fig. 1(d). These features have a width of only 3$\sim$5~meV, or a few percent of the total bandwidth of the UHB. This observation is enabled by the high energy resolution afforded by a lock-in technique using a bias modulation $V_{\textrm{mod}}$ = 1 mV or less, with a suitable sampling interval, and would be missed in measurements using the more typical modulation and sampling interval of $\sim$10 mV or more. We note that such features appear to be absent in the occupied states, and the reasons for this will be discussed below.

\begin{figure}
\centering
\includegraphics[scale=1]{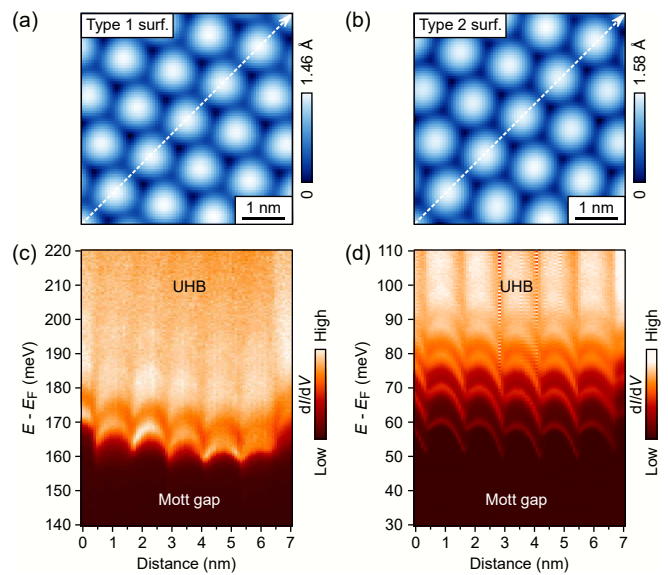}
\caption{\label{fig:2} Spectroscopic imaging at the UHB onset at Type 1 and 2 surfaces. (a) and (b) Constant-current topographs of each surface (setpoints $V$ = 0.25~V, $I$ = 1~nA, and $V$ = 0.11~V, $I$ = 0.44~nA, respectively). (c) and (d) Spectroscopic linecuts taken along the white dashed lines shown in (a) and (b). Note that each energy range is chosen for suitability to each UHB onset energy. From here onward, this work focuses on the features observed at the Type 2 surface.}
\end{figure}

Spatially-resolved tunneling spectroscopy measurements shown in Fig. 2 give a more complete overview of the behavior seen at the UHB edge at each surface. Figures 2(a) and (b) show topographs acquired simultaneously with spectroscopy measurements, and in these the Type 1 \& 2 surfaces appear nearly indistinguishable. Figures 2(c) and (d) show spectroscopic linecuts along each of the arrows shown in (a) and (b). Here the narrow peaks appear as arc-like features, with at least one arc (and five arcs) spanning each cluster for the Type 1 (Type 2) surface. This arc-like shape is consistent with a distortion by tip-induced effects of an intrinsically flat, tile-like structure. Despite this, the energy spacing between arcs appears rigid and position-independent. In the Supplementary Information, we further discuss the tip-induced effects, the underlying spatial distribution of the features as it relates to Mott localization, and also the differences between the Type 1 \& 2 spectra.

Below we examine the series of peaks at the Type 2 surface in particular, following the hypothesis that they result from replication of the zeroth peak by \textit{e-ph} interactions. An example of a typical spectrum acquired on a cluster at the Type 2 surface is analyzed, as shown in Fig. 3. Similar analyses of spectra acquired in different sample locations are shown in the Supplementary Information.

Given some sharp spectroscopic feature, the relative intensities of its \textit{e-ph} replicas are governed according to the Franck-Condon principle, which is commonly used to describe absorption line-shapes corresponding to an electronic excitation -- for example a photo-excited vertical transition -- coupled to a single optical phonon or vibron mode \cite{Ishizaka2008}. The Franck-Condon principle has also been used to interpret STM observations of the vibronic properties of organic molecules resting on an insulating substrate \cite{Wu2004,Qiu2004,Nazin2005}. There, injection of an electron into the lowest unoccupied molecular orbital (LUMO) formed a long-lived transient charged state, exciting molecular vibron modes which produced a series of replicas of the LUMO. Drawing an analogy with this, and regarding the SD cluster here as a `pseudo-molecule', the sharp state near the onset of the UHB can be thought of as corresponding to the LUMO. Instead of vibron modes, a charged state of the SD cluster may couple to lattice phonons. Generally, \textit{e-ph} replicas are discernible only when the width of the spectroscopic feature is small compared to the phonon energy. Where in previous cases, such as doped diamond \cite{Ishizaka2008} or the aforementioned molecular adsorbate, phonon or vibron energies were sufficient that replicas were seen for a broad band or molecular orbital, here only the narrow feature at the UHB edge is discernibly replicated. We return to the possible origin of the narrow state itself below.

The Franck-Condon scheme gives the relative intensities of replica peaks resulting from the emission of $n$ phonons. The intensity of the $n^{\textrm{th}}$ peak is
\begin{equation}
I_n = A e^{-D}\frac{1}{n!}D^n,
\end{equation}
where $A$ is a scaling coefficient common to the whole series, and $D$ is the Huang-Rhys parameter characterizing the strength of the electron-ion coupling \cite{MITOpenCourseWare}.

The expected line-shape for a conductance spectrum $g(V) \equiv \frac{\textrm{d}I}{\textrm{d}V}(V)$ can be expressed as a Franck-Condon progression of peaks, each convolved with a suitable broadening function, on a background function which in this case represents the UHB continuum. This can be written as
\begin{equation}
\begin{split}
g(V) & \propto \sum_{n=0}^{\infty} \bigg[ e^{-D}\frac{1}{n!}D^n \int \frac{ \Gamma_n / 2}{(q_e V)^2 +(\Gamma_n / 2)^2} \\ & \cdot \delta(q_e V - E_0 - nE_{\textrm{ph}})\ dV \bigg] \ + \textrm{UHB cont.,}
\end{split}
\end{equation}
where $n \in \{0, 1, 2... \}$, $\Gamma_n$ are peak-specific broadening parameters, $q_e$ is the electron charge, $E_0$ is the energy of the underlying narrow electronic state, and $E_{\textrm{ph}}$ is the phonon energy. The resulting ideal lineshapes given by Eqn. 2, ignoring the UHB continuum, are depicted for a few values of $D$ in Fig. 3(a).

In practice, in order to perform fitting to the conductance spectra and extract the key quantities $E_{\textrm{ph}}$, $D$ and $\Gamma_n$, a modified expression was used. First, $n$ was only allowed to run up to a value of four (for a total of five peaks), and in order to achieve convergence the intensities, energies, and widths of each of the individual peaks were treated as free parameters (marked with asterisks). Finally, convolution with the semicircular resolution function $\lambda(V)$ associated with the lock-in technique was taken into account:
\begin{equation}
g(V) \approx \sum_{n=0}^{4}\left [ \frac{ I_n^{*} \Gamma_n^{*} / 2}{(q_e V - E_n^{*})^2 +(\Gamma_n^{*} / 2)^2} * \lambda(V) \right ] \ + \textrm{UHB cont.,}
\end{equation}
where
\begin{equation}
\lambda(V) = 
\begin{cases}
\frac{1}{\pi V_{\textrm{mod}}} \sqrt{1-(\frac{V}{V_{\textrm{mod}}})^2} & \text{if } |V| \leq V_{\textrm{mod}}\\
0 & \text{otherwise.}
\end{cases}
\end{equation}
The result of fitting is shown in Fig. 3(b). A phenomenologically-driven choice was made for a line-shape to represent the UHB continuum, namely the upper half of a skew-normal function. Error bars shown in each plot correspond to the square roots of the diagonal elements of the estimated covariance matrix obtained alongside the optimized values.

\begin{figure}
\centering
\includegraphics[scale=1]{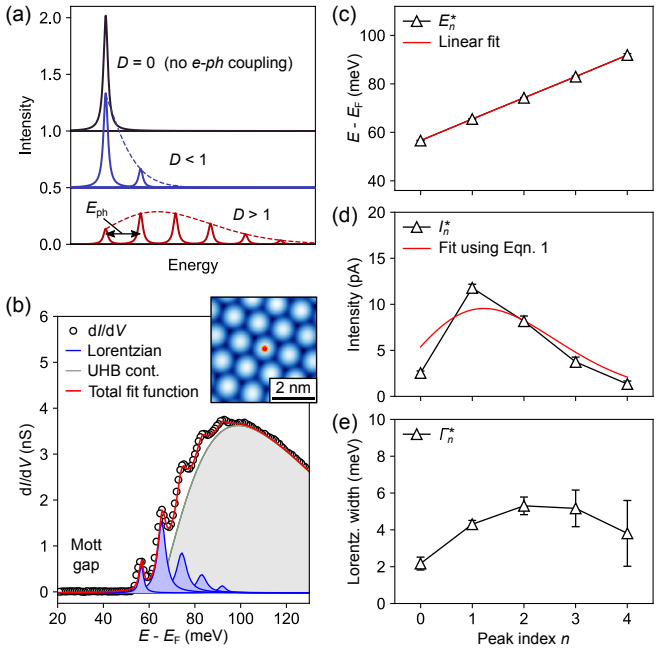}
\caption{\label{fig:3} Analysis of \textit{e-ph} replica peaks at the Type 2 surface. (a) Idealized lineshapes described by Eqn. 2 for various values of $D$. The dashed lines show envelope functions for the relative peak intensities according to the Franck-Condon principle (Eqn. 1). Plots are vertically offset for clarity. (b) Fitting, using Eqn. 3, to the conductance spectrum of unoccupied states at the Type 2 surface, acquired at the point shown in the inset. (c) Fitted energies for the peaks in (b), with a linear fit giving an average spacing of 8.8~meV. (d) Fitted peak intensities (total area under each Lorenztian curve), with further fitting using Eqn. 1 yielding the Huang-Rhys parameter $D$ = 1.75. (\textbf{e}) Fitted Lorentzian broadening parameters for each peak.}
\end{figure}

The resulting energies $E_n^{*}$ are plotted in Fig. 3(c), along with a linear fit that finally yields the average spacing $E_{\textrm{ph}}$ = 8.8~meV. Fitting to curves acquired in multiple locations on the sample shows that values of $E_{\textrm{ph}} \approx$ 9~meV are typical. 

The fitted peak intensities $I_n^{*}$ are plotted in Fig. 3(d), with the subsequently fitted curve for the Franck-Condon weights (from Eqn. 1) giving $D$ = 1.75. From the discussion which follows, we can infer only that this value is below the threshold for formation of a self-trapped polaron \cite{Stoneham2007}. The broadening of each peak, plotted in Fig. 3(e), shows the expected tendency that $\Gamma_n^{*}$ increases with higher energy, suggesting increased damping further into the UHB continuum. Here the most significant value is the broadening of the lowest-lying peak, $\Gamma_0^{*}$ = 2.17~meV, to which we will return below.

Figure 4 shows conductance spectra, and the result of fitting using Eqn. 3, obtained on a different Type 2 surface in a similar way to those shown in Fig. 3, for various tip-sample separations $z = z_{\mathrm{setpoint}} + z_{\mathrm{offset}}$. Varying $z$ has the effect of changing the electric field between the tip and sample (which is likely the source of the small energy shift with varying $z$), as well as changing the current by about one order of magnitude per 100 pm. The current is inversely proportional to the average time interval between tunneling events (see the discussion below). We see that $E_{\textrm{ph}}$, $D$ and $\Gamma_{0}^{*}$ do not show any systematic variation with $z$. This insensitivity is expected, as the phonon energy, electron-phonon coupling strength and peak width should be intrinsic properties of the sample.

\begin{figure}
\centering
\includegraphics[scale=1]{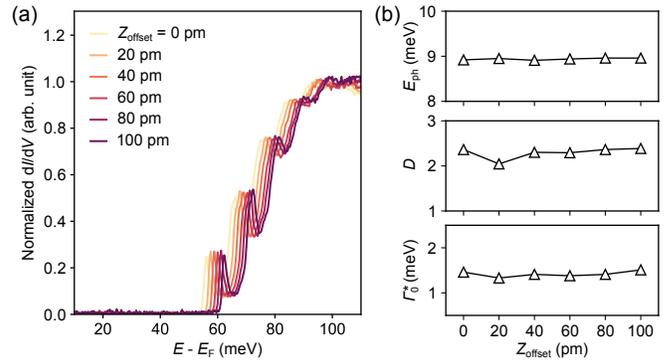}
\caption{\label{fig:4} Insensitivity of Franck-Condon series to tip-sample separation. (a) Conductance curves collected at varying tip STM tip height. After stabilizing the tip at the height $z_{\mathrm{setpoint}}$ (at $V$ = 0.11~V, $I$ = 0.44~nA), the tip height was altered by an amount $z_{\mathrm{offset}}$ before measuring the conductance spectrum. Here the curves are normalized for ease of comparison, using the value of $\mathrm{d}I/\mathrm{d}V$ at an energy 40 meV above that of the lowest-lying peak. (b) Results of the fitting with the function in Eqn. 3. The parameters $E_{\textrm{ph}}$, $D$ and $\Gamma_{0}^{*}$ are insensitive to the tip height.}
\end{figure}

\section{Discussion}
In order to identify the apparent phonon mode observed above, its energy can be compared against previously reported Raman, time-resolved photo-electron, and ultrafast optical spectroscopy measurements on the low temperature phase. The energy $E_{\textrm{ph}} \approx$ 9~meV corresponds to a Raman shift of about 71~cm$^{-1}$, and a strong Raman-active mode of this frequency has been observed, alongside other nearby modes \cite{Hangyo1983, Albertini2016}, though the phonon species were not identified. Time-resolved pump-probe spectroscopy experiments have observed photo-induced excitations coupling to the amplitude mode of the 1\textit{T}-TaS$_{2}$ CDW, at a frequency $f_{\textrm{AM}} \approx$ 2.4~THz (9.9~meV) \cite{Perfetti2008,Kusar2011,Stojchevska2014,Mann2016}. Some reports have also described a mode at a frequency of $f \approx$ 2.1~THz (8.6~meV) \cite{Toda2004, Dean2011}, consistent with one speculated as either an $E_{g}$ or $A_{1g}$ mode \cite{Albertini2016}. Like the amplitude mode, it appears only upon cooling into the commensurate CDW phase \cite{Albertini2016, Hu2018}.

From symmetry, it is reasonable that injection of an electron into the orbital localized at the SD center selectively excites the amplitude mode: The ionic arrangement in the host SD cluster responds to this excess negative charge by further symmetrically contracting towards the cluster center, setting the amplitude mode in motion. This scenario is depicted in Fig. 5(a). Though the observed phonon energy is also close to that of the previously reported $E_{g}$ or $A_{1g}$ mode, it is not clear how a phonon of such symmetry can be excited by injection of an electron into the centrally localized orbital of the cluster.

\begin{figure}
\centering
\includegraphics[scale=1]{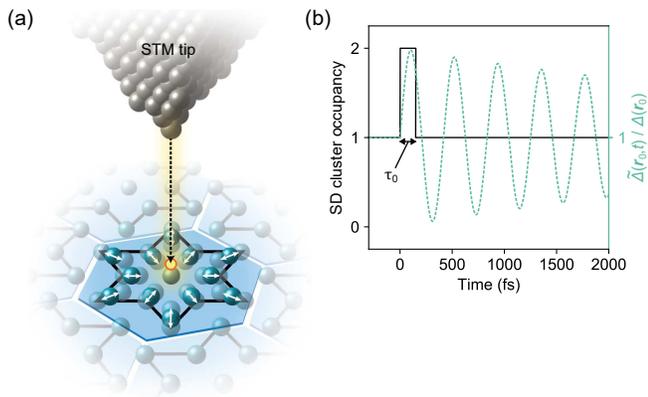}
\caption{\label{fig:5} Excitation of CDW amplitude mode by a transient doublon-like excitation. (a) A depiction of the excited amplitude mode in the SD cluster. (The Ta ionic displacements are greatly exaggerated.) (b) A qualitative plot of the SD cluster occupancy as a function of time, and the resulting local fluctuation of the order parameter.}
\end{figure}

Assuming that the orbital is fairly well localized to each cluster, the injected electron should be expected to dwell there for a finite time, before hopping away to neighboring clusters. This lifetime can be roughly estimated by considering the width of the zeroth peak in the series, $\Gamma_0^{*}$, and using the uncertainty relation: $\Gamma_{0}^{*} \tau_0 \geq \hbar / 2$. This yields a lifetime of $\tau _0 \approx$  152~fs for the spectrum shown in Fig. 3(b). (For comparison, the time-scale of electron hopping and screening in generic non-interacting systems is of the order 1~fs.) This long lifetime may attest to strong correlations in the half-filled background, since in the Fermi-Hubbard model the lifetime of a charged excitation residing in the UHB (a doublon) scales exponentially with the Mottness ratio (the ratio of the energy cost of double-occupancy $U$ to the bandwidth $W$) \cite{Sensarma2010}.

The time-scale for one period of the amplitude mode oscillation is $(1 / f_{\textrm{AM}}) \approx$ 417~fs. Figure 5(b) qualitatively shows the double-occupancy of the SD cluster as a transient perturbation which induces a response of the lattice. Taking the periodic lattice distortion (or equivalently the ionic charge modulation) as the order parameter, and writing it as $\psi(\textbf{\textit{Q}}) = \Delta(\textbf{\textit{Q}}) e^{i \phi}$, after a Fourier transform to real-space the amplitude can be sampled at the SD cluster center, $\textbf{\textit{r}}_{0}$, and labeled as $\Delta(\textbf{\textit{r}}_{0})$. This is the equilibrium about which the amplitude mode will oscillate ($\phi$ is fixed). Introducing time-dependence with $\psi(\textbf{\textit{r}},t)$, the modulated amplitude at $\textbf{\textit{r}}_{0}$ is $\tilde{\Delta}(\textbf{\textit{r}}_{0},t)$. The local modulation $\tilde{\Delta}(\textbf{\textit{r}}_{0},t) / \Delta (\textbf{\textit{r}}_{0})$ is qualitatively plotted as a function of time in Fig. 5(b) (green dashed curve) along with the cluster occupancy. Previously the amplitude mode has been observed to persist for a timescale on the order $\sim$10~ps \cite{Stojchevska2014}. Note that all the timescales depicted in Fig. 5(b), including the attenuation over time of the amplitude mode, are short compared to the average interval between tunneling of electrons from the tip to the sample. This time interval is $\sim$16~ns at the current of $\sim$10~pA that was typically recorded just above the zeroth peak for the setpoint parameters used for the measurements described in Figs. 2 \& 3. As the system recovers to equilibrium in a time much shorter than this interval, the phenomena described can be considered as the response of the system upon injection of a single electron.

From Fig. 5 it is reasonable to intuit that the lifetime $\tau_{0}$ $\approx$ 152~fs is of a suitable scale to trigger an oscillation of period $\approx$417~fs, since it covers most of the oscillation's first upward swing. Corollary to this, if $\tau_{0}$ tended toward zero, so the coupling to the oscillation would diminish, i.e. $D$ would tend to zero. In the opposite limit, $\tau_{0} \rightarrow \infty$ (self-trapped polaron), the amplitude mode excitation would still be expected, oscillating around a new equilibrium lattice distortion. 

We now return to the origin of the peaks observed at both the Type 1 \& 2 surfaces. The extraordinary narrowness of the features, as well as the \textit{a priori} condition of strong electronic correlations, likely precludes any conventional band-theory explanation, such as a flat band or other source of a van Hove singularity.

A possible route to an explanation is provided by dynamical mean field theory (DMFT) \cite{Georges1996}. Sharp structures at the Hubbard band edges have been found in treatments of both the dimer \cite{Najera2017,Najera2018} and single-site Hubbard models \cite{Nishimoto2004, Granath2014}. These seem to explain such aspects as the lifetime for double-occupancy at the quasiparticle peak (analogous to $\tau_{\textrm{0}}$) \cite{Ganahl2015}, and the asymmetry between features observed in the UHB and LHB \cite{Granath2014, Najera2018}. However, the dimer model is not applicable to the Type 2 (un-dimerized) surface, and though the single-site model is, the existence of quasiparticles in that model has yet to be firmly established.

The salience of \textit{e-ph} coupling in the real material (absent in DMFT treatments mentioned above) suggests an alternative, intuitive interpretation for the narrow peaks as polaronic bound states. Excess electrons injected into a Mott insulator (doublons) can in principle interact with other excitations to form bound states (see for example \cite{Terashige2019}). In this case, the charge of the additional electron may be partially screened by the ionic displacement associated with lattice distortions (phonons), reducing the on-site repulsion $U$ and allowing the bound state to split off from the UHB. This interpretation is supported by the fact that the lowest-lying peak at the Type 2 surface appears to reside outside the UHB, rather than within it.

As a final observation, in contrast to the clear and abrupt onset of the UHB, the soft and poorly defined onset of occupied states at each type of surface [seen in Fig. 1(d)], which likely results from the hole-doped character of the sample, may explain the apparent absence of narrow structures at the LHB edge, which would correspond to creation of holons by tunneling of electrons from the sample to the tip, and their respective coupling with the amplitude mode. Markers of such interactions, if they exist, may be smeared out or small enough that they are concealed by noise.

In summary, we have observed dynamics induced by electron tunneling into the UHB of Mott insulating 1\textit{T}-TaS$_{2}$ surfaces. This reveals unusual, narrow spectral features possibly corresponding to long-lived quasiparticles or bound states, which couple to the amplitude mode of the CDW, the precursor to the Mott state. These observations provide a nano-scale microscopic view of non-equilibrium behavior, complementary to those observations previously provided using ultra-fast optical and time-resolved photo-emission spectroscopy techniques. Furthermore, they lay down a challenge for theories supporting quantitative model calculations to understand the origin of the unusual UHB edge features, which we anticipate will yield as-yet unknown physics of Mott-Hubbard systems more generally.

\section{Materials and Methods}
Crystals of 1\textit{T}-TaS$_{2}$ were synthesized and prepared by cleaving at $\approx$77~K in UHV ($P \sim 10^{-10}$ Torr) as described previously \cite{Tani1981, Butler2020}, before insertion into a modified Unisoku 1300 low-temperature STM system held at 1.5~K \cite{Hanaguri2006}. STM measurements were performed using electro-chemically etched tungsten tips, which were characterized and fine-tuned using field ion microscopy and mild indentation at a clean Cu(111) surface. $\frac{\textrm{d}I}{\textrm{d}V}$ conductance was measured using the lock-in technique with frequency $f_{\textrm{mod}}$ = 617.3~Hz and bias modulation of amplitude $V_{\textrm{mod}}$ = 1~mV.

\section*{Acknowledgements}
We are grateful to Y. Kohsaka, T. Machida, P. A. Lee, D. Mihailovic, Y. Nomura, N. Nagaosa, M. Civelli, V. Dobrosavljevi\'{c} and M. J. Rozenberg for helpful discussions. C.J.B. acknowledges support from RIKEN's Programs for Junior Scientists. This work was supported in part by JSPS KAKENHI grant numbers JP18K13511, JP19H00653, JP19H01855 and JP19H05602.

\subsection*{Author contributions}
C.J.B, T.H. and Y.I. conceived the project, and M.Y. and Y.I. synthesized the 1\textit{T}-TaS$_{2}$ crystals. C.J.B. performed the STM measurements, interpreted the data and prepared the manuscript with input from all authors.

\section*{Data availability}
The data that support the findings presented here are available from the corresponding authors upon reasonable request.


\begin{thebibliography}{99}


\bibitem{Wilson1975}
J. A. Wilson, F. J. Di Salvo, and S. Mahajan,
\textit{Charge-density waves and superlattices in the metallic layered transition metal dichalcogenides.}
Adv. Phys. \textbf{24,}  117--201 (1975).
\url{https://doi.org/10.1080/00018737500101391}


\bibitem{Johannes2008}
M. D. Johannes and I. I. Mazin
\textit{Fermi surface nesting and the origin of charge density waves in metals.}
Phys. Rev. B \textbf{77,} 165135 (2008).
\url{https://doi.org/10.1103/PhysRevB.77.165135}


\bibitem{Bardeen1957}
J. Bardeen, L. N. Cooper, and J. R. Schrieffer,
\textit{Theory of Superconductivity.}
Phys. Rev. \textbf{108,} 1175 (1957).
\url{https://doi.org/10.1103/PhysRev.108.1175}


\bibitem{Eliashberg1960}
G. M. Eliashberg,
\textit{Interactions between Electrons and Lattice Vibrations in a Superconductor.}
J. Exp. Theor. Phys. \textbf{11,} 696--702 (1960).
\url{http://www.jetp.ac.ru/cgi-bin/e/index/e/11/3/p696?a=list}


\bibitem{Fazekas1979}
P. Fazekas and E. Tosatti,
\textit{Electrical, structural and magnetic properties of pure and doped 1T-TaS$_{2}$.}
Phil. Mag. B \textbf{39,} 229--244 (1979).
\url{https://doi.org/10.1080/13642817908245359}


\bibitem{Fazekas1980}
P. Fazekas and E. Tosatti,
\textit{Charge carrier localization in pure and doped 1T-TaS$_{2}$.}
Physica B \& C \textbf{99,} 183--187 (1980).
\url{https://doi.org/10.1016/0378-4363(80)90229-6}


\bibitem{Perfetti2008}
L. Perfetti, P. A. Loukakos, M. Lisowski, U. Bovensiepen, M. Wolf, H. Berger, S. Biermann, and A. Georges,
\textit{Femtosecond dynamics of electronic states in the Mott insulator 1T-TaS$_{2}$ by time resolved photoelectron spectroscopy.}
New J. Phys. \textbf{10,} 053019 (2008).
\url{https://doi.org/10.1088/1367-2630/10/5/053019}


\bibitem{Mann2016}
A. Mann, E. Baldini, A. Odeh, A. Magrez, H. Berger, and F. Carbone,
\textit{Probing the coupling between a doublon excitation and the charge-density wave in TaS$_{2}$ by ultrafast optical spectroscopy.}
Phys. Rev. B \textbf{94,} 115122 (2016).
\url{https://doi.org/10.1103/PhysRevB.94.115122}


\bibitem{Ligges2018} 
M. Ligges, I. Avigo, D. Gole\v{z}, H. U. R. Strand, Y. Beyazit, K. Hanff, F. Diekmann, L. Stojchevska, M. Kall\"{a}ne, P. Zhou, K. Rossnagel, M. Eckstein, P. Werner, and U. Bovensiepen,
\textit{Ultrafast Doublon Dynamics in photoexcited 1\textit{T}-TaS$_{2}$.}
Phys. Rev. Lett. \textbf{120,} 166401 (2018).
\url{https://doi.org/10.1103/PhysRevLett.120.166401}


\bibitem{Petersen2011}
J. C. Petersen, S. Kaiser, N. Dean, A. Simoncig, H. Y. Liu, A. L. Cavalieri, C. Cacho, I. C. E. Turcu, E. Springate, F. Frassetto, L. Poletto, S. S. Dhesi, H. Berger, and A. Cavalleri,
\textit{Clocking the Melting Transition of Charge and Lattice Order in 1\textit{T}-TaS$_{2}$ with Ultrafast Extreme-Ultraviolet Angle-Resolved Photoemission Spectroscopy.}
Phys. Rev. Lett. \textbf{107,} 177402 (2011).
\url{https://doi.org/10.1103/PhysRevLett.107.177402}


\bibitem{Hellmann2012}
S. Hellmann, T. Rohwer, M. Kall\"{a}ne, K. Hanff, C. Sohrt, A. Stange, A. Carr, M.M. Murnane, H.C. Kapteyn, L. Kipp, M. Bauer, and K. Rossnagel,
\textit{Time-domain classification of charge-density-wave insulators.}
Nat. Commun. \textbf{3,} 1069 (2012).
\url{https://doi.org/10.1038/ncomms2078}


\bibitem{Zhang2019}
J. Zhang, C. Lian, M. Guan, W. Ma, H. Fu, H. Guo, and S. Meng,
\textit{Photoexcitation Induced Quantum Dynamics of Charge Density Wave and Emergence of a Collective Mode in 1\textit{T}-TaS$_{2}$.}
Nano Lett. \textbf{19,} (9) 6027–-6034 (2019).
\url{https://doi.org/10.1021/acs.nanolett.9b01865}


\bibitem{Avigo2019}
I. Avigo, F. Queisser, P. Zhou, M. Ligges, K. Rossnagel, R. Sch\"{u}tzhold, and U. Bovensiepen,
\textit{Doublon bottleneck in the ultrafast relaxation dynamics of hot electrons in 1\textit{T}-TaS$_{2}$.}
Phys. Rev. Res. \textbf{2,} 022046(R) (2019).
\url{https://doi.org/10.1103/PhysRevResearch.2.022046}


\bibitem{Ritschel2018}
T. Ritschel, H. Berger, and J. Geck,
\textit{Stacking-driven gap formation in layered 1\textit{T}-TaS$_{2}$.}
Phys. Rev. B \textbf{98,} 195134 (2018).
\url{https://doi.org/10.1103/PhysRevB.98.195134}


\bibitem{Lee2019}
S.-H. Lee, J. S. Goh, and D. Cho,
\textit{Origin of the Insulating Phase and First-Order Metal-Insulator Transition in 1\textit{T}-TaS$_{2}$.}
Phys. Rev. Lett. \textbf{122,} 106404 (2019).
\url{https://doi.org/10.1103/PhysRevLett.122.106404}


\bibitem{Tanda1984}
S. Tanda, T. Sambongi, T. Tani, and S. Tanaka,
\textit{X-Ray Study of Charge Density Wave Structure of 1T-TaS$_{2}$.}
J. Phys. Soc. Jpn. \textbf{53,} 476--479 (1984).
\url{https://doi.org/10.1143/JPSJ.53.476}


\bibitem{Naito1984}
M. Naito, H. Nishihara, and S. Tanaka,
\textit{Nuclear quadrupole resonance in the charge density wave state of 1\textit{T}-TaS$_{2}$.}
J. Phys. Soc. Jpn. \textbf{53,} 1610--1613 (1984).
\url{https://doi.org/10.1143/JPSJ.53.1610}


\bibitem{Naito1986}
M. Naito, H. Nishihara, and S. Tanaka,
\textit{Nuclear magnetic resonance and nuclear quadrupole resonance study of $^{181}$Ta in the commensurate charge density wave state of 1\textit{T}-TaS$_{2}$.}
J. Phys. Soc. Jpn. \textbf{55,} 2410--2421 (1986).
\url{https://doi.org/10.1143/JPSJ.55.2410}


\bibitem{Stahl2019}
Q. Stahl, M. Kusch, F. Heinsch, G. Garbarino, N. Kretzschmar, K. Hanff, K. Rossnagel, J. Geck, and T. Ritschel,
\textit{Collapse of layer dimerization in the photo-induced hidden state of 1T-TaS$_{2}$.}
Nat. Commun. \textbf{11,} 1247 (2020).
\url{https://doi.org/10.1038/s41467-020-15079-1}


\bibitem{LeGuyader2017}
L. Le Guyader, T. Chase, A. H. Reid, R. K. Li, D. Svetin, X. Shen, T. Vecchione, X. J. Wang, D. Mihailovic, and H. A. D\"{u}rr,
\textit{Stacking order dynamics in the quasi-two-dimensional dichalcogenide 1\textit{T}-TaS$_{2}$ probed with MeV ultrafast electron diffraction.}
Struct. Dyn. \textbf{4,} 044020 (2017).
\url{https://doi.org/10.1063/1.4982918}


\bibitem{vonWitte2019}
G. von Witte, T. Ki{\ss}linger, J. G. Horstmann, K. Rossnagel, M. A. Schneider, C. Ropers, and L. Hammer,
\textit{Surface structure and stacking of the commensurate $(\sqrt{13} \times \sqrt{13})R13.9 ^{\circ}$ charge density wave phase of 1\textit{T}-TaS$_{2}$(0001).}
Phys. Rev. B \textbf{100,} 155407 (2019).
\url{https://doi.org/10.1103/PhysRevB.100.155407}


\bibitem{Butler2020}
C. J. Butler, M. Yoshida, T. Hanaguri, and Y. Iwasa,
\textit{Mottness versus unit-cell doubling as the driver of the insulating state in 1\textit{T}-TaS$_{2}$.}
Nat. Commun. \textbf{11,} 2477 (2020).
\url{https://doi.org/10.1038/s41467-020-16132-9}


\bibitem{Wang2020}
Y. D. Wang, W. L. Yao, Z. M. Xin, T. T. Han, Z. G. Wang, L. Chen, C. Cai, Yuan Li, and Y. Zhang,
\textit{Band insulator to Mott insulator transition in 1\textit{T}-TaS$_{2}$.}
Nat. Commun. \textbf{11,} 4215 (2020).
\url{https://doi.org/10.1038/s41467-020-18040-4}


\bibitem{Fuhrmann2006}
A. Fuhrmann, D. Heilmann, and H. Monien,
\textit{From Mott insulator to band insulator: a dynamical mean-field theory study. }
Phys. Rev. B \textbf{73,} 245118 (2006).
\url{https://doi.org/10.1103/PhysRevB.73.245118}


\bibitem{Bu2019}
K. Bu, W. Zhang, Y. Fei, Z. Wu, Y. Zheng, J. Gao, X. Luo, Y.-P. Sun, and Y. Yin,
\textit{Possible strain induced Mott gap collapse in 1\textit{T}-TaS$_{2}$.}
Commun. Phys. \textbf{2,} 146 (2019). 
\url{https://doi.org/10.1038/s42005-019-0247-0}


\bibitem{Ishizaka2008}
K. Ishizaka, R. Eguchi, S. Tsuda, A. Chainani, T. Yokoya, T. Kiss, T. Shimojima, T. Togashi, S. Watanabe, C.-T. Chen, Y. Takano, M. Nagao, I. Sakaguchi, T. Takenouchi, H. Kawarada, and S. Shin,
\textit{Temperature-Dependent Localized Excitations of Doped Carriers in Superconducting Diamond.}
Phys. Rev. Lett. \textbf{100,} 166402 (2008).
\url{https://doi.org/10.1103/PhysRevLett.100.166402}


\bibitem{Wu2004}
S. W. Wu, G. V. Nazin, X. Chen, X. H. Qiu, and W. Ho,
\textit{Control of Relative Tunneling Rates in Single Molecule Bipolar Electron Transport.}
Phys. Rev. Lett. \textbf{93,} 236802 (2004).
\url{https://doi.org/10.1103/PhysRevLett.93.236802}


\bibitem{Qiu2004}
X. H. Qiu, G. V. Nazin, and W. Ho,
\textit{Vibronic States in Single Molecule Electron Transport.}
Phys. Rev. Lett. \textbf{92,} 206102 (2004).
\url{https://doi.org/10.1103/PhysRevLett.92.206102}


\bibitem{Nazin2005}
G. V. Nazin, S. W. Wu, and W. Ho,
\textit{Tunneling rates in electron transport through double-barrier molecular junctions in a scanning tunneling microscope.}
Proc. Natl. Acad. Sci. U.S.A. \textbf{102} (25), 8832--8837 (2005).
\url{https://doi.org/10.1073/pnas.0501171102}


\bibitem{MITOpenCourseWare}
A. Tokmakoff,
Introductory Quantum Mechanics II,
Spring 2009,
Massachusetts Institute of Technology,
Cambridge~MA,
MIT OpenCourseWare,
\url{https://ocw.mit.edu/courses/chemistry/5-74-introductory-quantum-mechanics-ii-spring-2009/lecture-notes/MIT5_74s09_lec08.pdf}
License: Creative Commons BY-NC-SA.


\bibitem{Stoneham2007}
A. M. Stoneham, J. Gavartin, A. L. Shluger, A. V. Kimmel,
D. Mu\~{n}oz Ramo, H. M. R{\o}nnow, G. Aeppli, and C. Renner,
\textit{Trapping, self-trapping and the polaron family.}
J. Phys.: Condens. Matter \textbf{19,} 255208 (2007).
\url{https://doi.org/10.1088/0953-8984/19/25/255208}


\bibitem{Hangyo1983}
M. Hangyo, S.-I. Nakashima, and A. Mitsuishi,
\textit{Raman spectroscopic studies of MX2-type layered compounds.}
Ferroelectrics \textbf{52,} 151--159 (1983).
\url{https://doi.org/10.1080/00150198308208248}


\bibitem{Albertini2016}
O. R. Albertini, R. Zhao, R. L. McCann, S. Feng, M. Terrones, J. K. Freericks, J. A. Robinson, and A. Y. Liu,
\textit{Zone-center phonons of bulk, few-layer, and monolayer 1\textit{T}-TaS$_{2}$: Detection of commensurate charge density wave phase through Raman scattering.}
Phys. Rev. B \textbf{93,} 214109 (2016).
\url{https://doi.org/10.1103/PhysRevB.93.214109}


\bibitem{Kusar2011}
P. Kusar, T. Mertelj, V. V. Kabanov, J.-H. Chu, I. R. Fisher, H. Berger, L. Forr\'{o}, and D. Mihailovic,
\textit{Anharmonic order-parameter oscillations and lattice coupling in strongly driven 1T-TaS$_{2}$ and TbTe$_{3}$ charge-density-wave compounds: A multiple-pulse femtosecond laser spectroscopy study.}
Phys. Rev. B \textbf{83,} 035104 (2011).
\url{https://doi.org/10.1103/PhysRevB.83.035104}


\bibitem{Stojchevska2014}
L. Stojchevska, I. Vaskivskyi, T. Mertelj, P. Kusar, D. Svetin, S. Brazovskii, and D. Mihailovic,
\textit{Ultrafast Switching to a Stable Hidden Quantum State in an Electronic Crystal.}
Science \textbf{344,} 177--180 (2014).
\url{https://doi.org/10.1126/science.1241591}


\bibitem{Toda2004}
Y. Toda, K. Tateishi, and S. Tanda,
\textit{Anomalous coherent phonon oscillations in the commensurate phase of the quasi-two-dimensional 1\textit{T}-TaS$_{2}$ compound.}
Phys. Rev. B \textbf{70,} 033106 (2004).
\url{https://doi.org/10.1103/PhysRevB.70.033106}


\bibitem{Dean2011}
N. Dean, J. C. Petersen, D. Fausti, R. I. Tobey, S. Kaiser, L. V. Gasparov, H. Berger, and A. Cavalleri,
\textit{Polaronic Conductivity in the Photoinduced Phase of 1\textit{T}-TaS$_{2}$.}
Phys. Rev. Lett. \textbf{106,} 016401 (2011).
\url{https://doi.org/10.1103/PhysRevLett.106.016401}


\bibitem{Hu2018}
Q. Hu, C. Yin, L. Zhang, L. Lei, Z. Wang, Z. Chen, J. Tang, and R. Ang,
\textit{Direct observation of melted Mott state evidenced from Raman scattering in 1T-TaS$_{2}$ single crystal.}
Chinese Phys. B \textbf{27,} 017104 (2018).
\url{https://doi.org/10.1088/1674-1056/27/1/017104}


\bibitem{Sensarma2010}
R. Sensarma, D. Pekker, E. Altman, E. Demler, N. Strohmaier, D. Greif, R. J\"{o}rdens, L. Tarruell, H. Moritz, and T. Esslinger,
\textit{Lifetime of double occupancies in the Fermi-Hubbard model.}
Phys. Rev. B \textbf{82,} 224302 (2010).
\url{https://doi.org/10.1103/PhysRevB.82.224302}


\bibitem{Georges1996}
A. Georges, G. Kotliar, W. Krauth, and M. J. Rozenberg,
\textit{Dynamical mean-field theory of strongly correlated fermion systems and the limit of infinite dimensions.}
Rev. Mod. Phys. \textbf{68,} 13 (1996).
\url{https://doi.org/10.1103/RevModPhys.68.13}


\bibitem{Najera2017}
O. N\'{a}jera, M. Civelli, V. Dobrosavljevi\'{c}, and M. J. Rozenberg,
\textit{Resolving the VO$_{2}$ controversy: Mott mechanism dominates the insulator-to-metal transition.}
Phys. Rev. B \textbf{95,} 035113 (2017).
\url{https://doi.org/10.1103/PhysRevB.95.035113}


\bibitem{Najera2018}
O. N\'{a}jera, M. Civelli, V. Dobrosavljevi\'{c}, and M. J. Rozenberg,
\textit{Multiple crossovers and coherent states in a Mott-Peierls insulator.}
Phys. Rev. B \textbf{97,} 045108 (2018).
\url{https://doi.org/10.1103/PhysRevB.97.045108}


\bibitem{Nishimoto2004}
S. Nishimoto, F. Gebhard, and E. Jeckelmann,
\textit{Dynamical density-matrix renormalization group for the Mott–Hubbard insulator in high dimensions.}
J. Phys.: Condens. Matter \textbf{16,} 7063 (2004).
\url{https://doi.org/10.1088/0953-8984/16/39/038}


\bibitem{Granath2014}
M. Granath, and J. Sch\"{o}tt,
\textit{Signatures of coherent electronic quasiparticles in the paramagnetic Mott insulator.}
Phys. Rev. B \textbf{90,} 235129 (2014).
\url{https://doi.org/10.1103/PhysRevB.90.235129}


\bibitem{Ganahl2015}
M. Ganahl, M. Aichhorn, H. G. Evertz, P. Thunstr\"{o}m, K. Held, and F. Verstraete,
\textit{Efficient DMFT impurity solver using real-time dynamics with matrix product states.}
Phys. Rev. B \textbf{92,} 155132 (2015).
\url{https://doi.org/10.1103/PhysRevB.92.155132}


\bibitem{Terashige2019}
T. Terashige, T. Ono, T. Miyamoto, T. Morimoto, H. Yamakawa, N. Kida, T. Ito, T. Sasagawa, T. Tohyama, and H. Okamoto,
\textit{Doublon-holon pairing mechanism via exchange interaction in two-dimensional cuprate Mott insulators. }
Sci. Adv. \textbf{5,} eaav2187 (2019).
\url{https://doi.org/10.1126/sciadv.aav2187}


\bibitem{Tani1981}
T. Tani, K. Okajima, T. Itoh, and S. Tanaka,
\textit{Electronic transport properties in 1T-TaS$_{2}$.}
Physica B \& C \textbf{105,} 127--131 (1981).
\url{https://doi.org/10.1016/0378-4363(81)90230-8}


\bibitem{Hanaguri2006}
T. Hanaguri,
\textit{Development of high-field STM and its application to the study on magnetically tuned criticality in Sr$_{3}$Ru$_{2}$O$_{7}$.}
J. Phys. Conf. Ser. \textbf{51,} 514 (2006).
\url{https://doi.org/10.1088/1742-6596/51/1/117}


\end{thebibliography}
\end{document}


\title{Supplementary Information for: \protect\\
``Doublon-like excitations and their phononic coupling \protect\\ in a Mott charge-density-wave system''}

\author{C. J. Butler}
\email{christopher.butler@riken.jp}
\affiliation{RIKEN Center for Emergent Matter Science, 2-1 Hirosawa, Wako, Saitama 351-0198, Japan}

\author{M. Yoshida}
\affiliation{RIKEN Center for Emergent Matter Science, 2-1 Hirosawa, Wako, Saitama 351-0198, Japan}

\author{T. Hanaguri}
\email{hanaguri@riken.jp}
\affiliation{RIKEN Center for Emergent Matter Science, 2-1 Hirosawa, Wako, Saitama 351-0198, Japan}

\author{Y. Iwasa}
\affiliation{RIKEN Center for Emergent Matter Science, 2-1 Hirosawa, Wako, Saitama 351-0198, Japan}
\affiliation{Quantum-Phase Electronics Center and Department of Applied Physics, The University of Tokyo, 7-3-1 Hongo, Bunkyo-ku, Tokyo 113-8656, Japan}

\maketitle

\section{Setpoint-related and tip-induced effects.}
\subsection{Interpretation of the topographs in Fig. 2.}
To aid the following discussion of tip-induced effects, and the interpretation of spectroscopic imaging data shown in Fig. 2 of the main work and Figs. S1 \& S3 here, it is worth noting what the accompanying topography maps represent. It is well-known that in the constant-current mode, scanning tunneling microscopy (STM) topographs generally do not simply represent the height of atoms at the sample surface, but rather surfaces of constant integrated local density of states of the sample $\rho_{\mathrm{samp.}}$, according to the commonly adopted approximation 
\begin{equation*}
I \approx -\frac{4 \pi q_{e}}{\hbar} |M|^{2} \rho_{\mathrm{tip}} \int_{E_{\mathrm{F,samp.}}}^{q_{e}V} \rho_{\mathrm{samp.}}(\epsilon) d \epsilon,
\end{equation*}
in which $E_{\mathrm{F,samp.}} \equiv 0$ eV, and the energy dependence of the tunneling matrix elements $M$ and the tip density of states $\rho_{\mathrm{tip}}$ are assumed to be negligible. Here $q_{e}$ is the electronic charge and $V$ is the sample bias voltage.

In the specific case presented in Fig. 2(a) [and (b)] of the main text, the interval between $E_{\mathrm{F,samp.}}$ and $q_{e}V$ = 250~meV [and $q_{e}V$ = 110~meV] only includes the Mott-localized orbitals of the upper Hubbard band (UHB). The conduction band lies beyond this range. Hence, as a crude approximation, the topograph maps the surface of equal probability to tunnel into the Mott-localized Ta $5d_{z^{2}}$ orbital at the cluster center nearest to any given (\textit{x},\textit{y}) location of the tip. This can be thought of as the three-dimensional boundary of the `catchment volume' belonging to each cluster, somewhat analogous to a Voronoi partition. An important outcome of this is that for bias voltages up to a setpoint which only includes the UHB, electrons tunnel into the centrally Mott-localized orbital with equal probability (equal conductance) regardless of whether the tip is positioned above the cluster's center or towards its periphery. The spectroscopic imaging data presented in this work should be interpreted with this in mind.

\subsection{Characterization of tip-induced energy shifts.}
Below we give a more complete view of the spectroscopic imaging data shown in Fig. 2 of the main text.

\begin{figure}[!ht]
\centering
\includegraphics[scale=1]{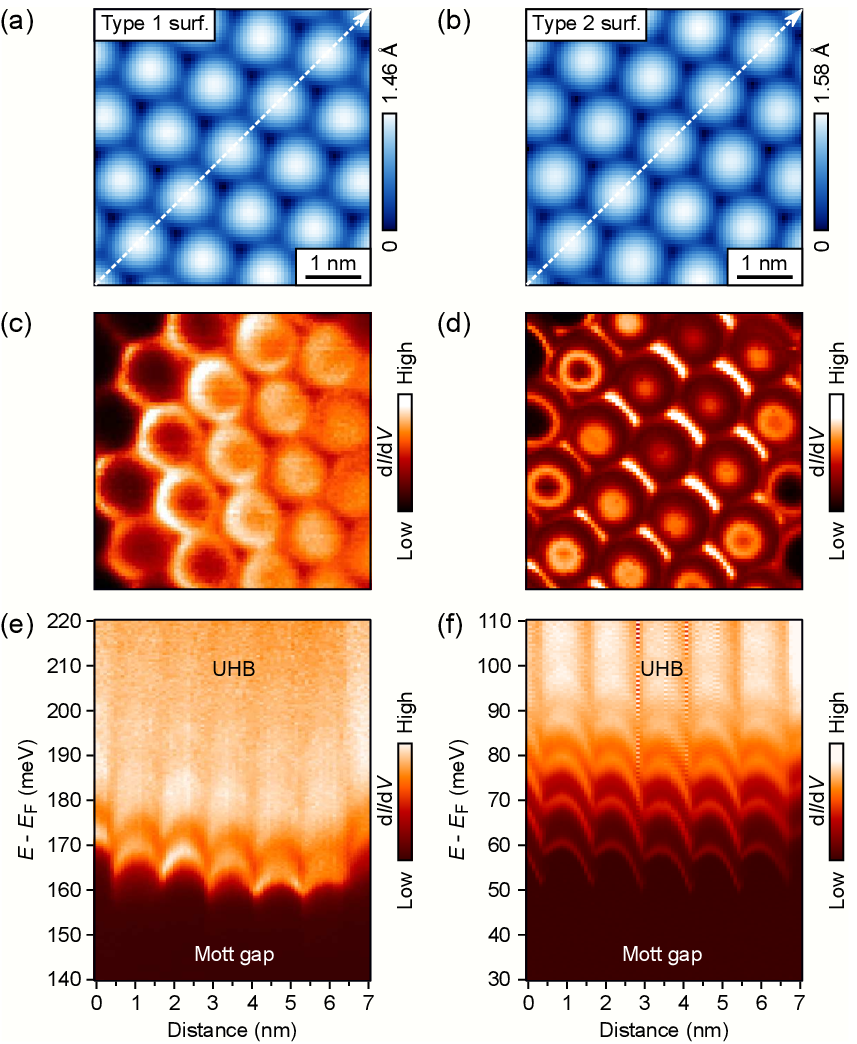}
\caption{\label{fig:1} Overview of tip-induced influence on spectroscopic imaging data. (a) and (b) Constant-current topographic images of each surface (setpoints \textit{V} = 0.25~V, \textit{I} = 1~nA, and \textit{V} = 0.11~V, \textit{I} = 0.44~nA, respectively). (c) and (d) Selected conductance maps, at 165~mV and 60~mV respectively. (e) and (f) Spectroscopic linecuts taken along the white dashed lines shown in (a) and (b).}
\end{figure}

Figures S1(a) and (b) reproduce the topographs shown in Fig. 2 of the main manuscript. Conductance maps in each respective field of view, at selected values of bias voltage, are shown in Figs. S1(c) and (d). It is important to recognize that these images do not represent the spatial distributions of the underlying peak features discussed in the main text, because they are significantly influenced by so-called tip-induced band bending (TIBB) effects. The complex ring-like features are amenable to interpretation in combination with the conductance linecuts [(e) and (f)] across the diagonal of each field of view [along each of the arrows shown in (a) and (b)]. The peaks appear in these linecuts as arc-like features, with one arc (and five arcs) spanning each cluster for the Type 1 (Type 2) surface. Given this arc-like appearance, it can be recognized that the ring-like features appearing in the conductance map in panel (c) and especially in (d), actually represent planar ($x,y$) cuts through dome-like structures in the ($x,y,E$) space. This dome-like shape is likely a distortion of an intrinsically flat, tile-like structure, due to TIBB effects.

Below we attempt to characterize and compensate for the TIBB effects in order to retrieve a more accurate image of the spatial distribution of the lowest-lying conductance peak feature. The strategy is to acquire spectra at various tip heights, observe the shifts of the lowest-lying peak, and characterize the trend. Finally we extend this characterization to the entire field-of-view to compensate for the TIBB in the spectroscopic imaging data.

Starting with the assumption that the sample surface is located in a flat plane we express the tip height above this plane as $z_{\mathrm{tip}}(\textbf{\textit{r}}) = z_{\mathrm{setpoint}}(\textbf{\textit{r}}) + z_{\mathrm{offset}}$, where the constant-current topograph gives $z_{\mathrm{setpoint}}(\textbf{\textit{r}})$. We then vary $z_{\mathrm{offset}}$, as well as exploiting the large variation within $z_{\mathrm{setpoint}}(\textbf{\textit{r}})$, in order to sample a wide range of tip heights. This is shown schematically in Fig. S2(a). For the Type 1 surface, the minimum height is $z_{\mathrm{setpoint}}(\triangledown)$, and the maximum is $z_{\mathrm{setpoint}}(\circ)$ + 120 pm, so that the measurements span a range of nearly 300~pm. (Although the absolute tip-sample separation cannot be known, it is usually estimated to be 600$\sim$800~pm in ordinary constant current feedback mode.)

\begin{figure}
\centering
\includegraphics[scale=1]{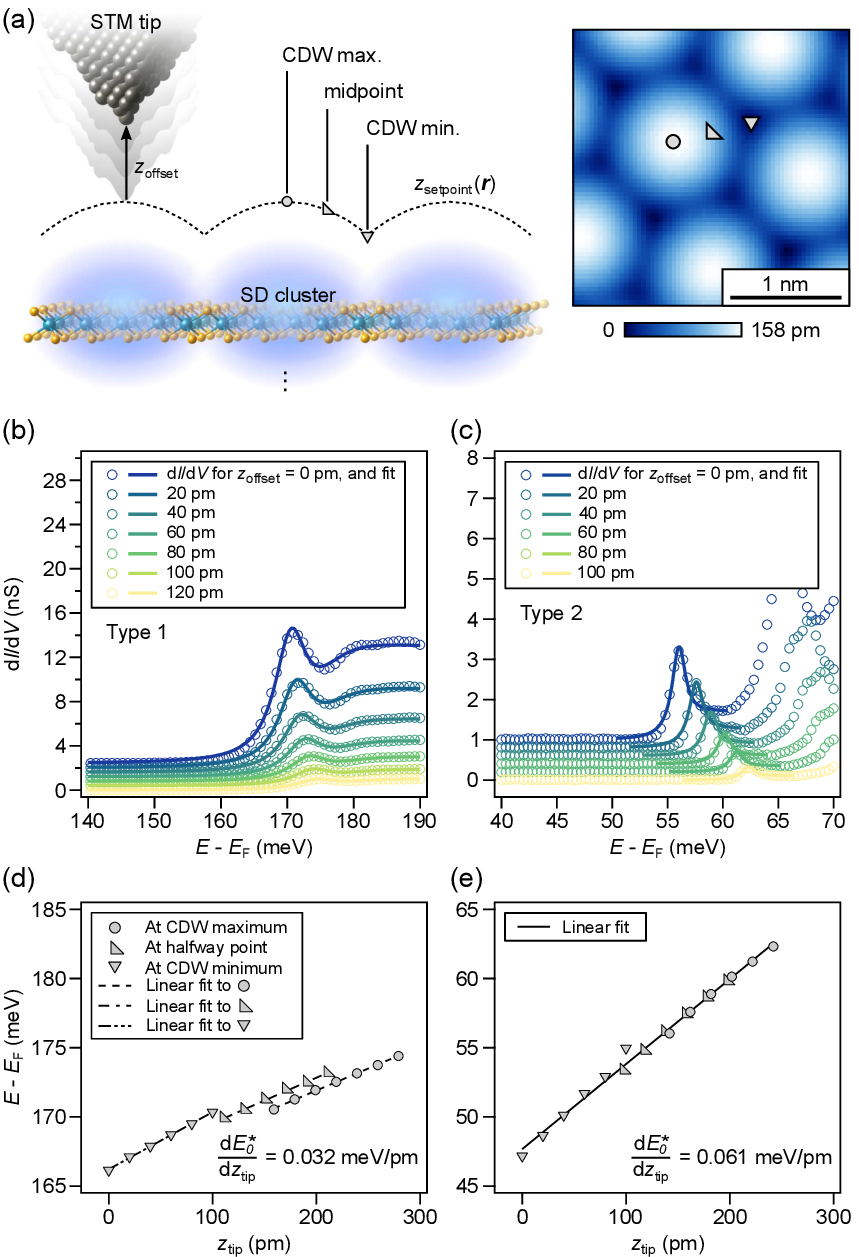}
\caption{\label{fig:2} Characterization of tip induced energy shift $\mathrm{d}E/\mathrm{d}z_{\mathrm{tip}}$ for each surface. (a) A schematic of the procedure for collecting $z_{\mathrm{offset}}$-dependent d\textit{I}/d\textit{V}(V) spectra. The locations of the CDW maximum and minimum apparent heights, and the location of the point halfway between are marked in the topographic image, with a circle, triangle and wedge, respectively. (The topograph shown is for the Type 2 surface. \textit{V} = 110~mV, \textit{I} = 0.44~nA.) (b) and (c) $\mathrm{d}I/\mathrm{d}V$ curves acquired at $z_{\mathrm{setpoint}}(\circ) + z_{\mathrm{offset}}$ for SD clusters at the Type 1 and 2 surfaces. Curves are vertically offset by 0.4 nS and 0.2 nS, respectively. (d) and (e) The fitted peak positions $E_0^{*}(z_{\mathrm{tip}})$ for the Type 1 and 2 surfaces, respectively. In (d), $\mathrm{d}E_{0}^{*}/\mathrm{d}z_{\mathrm{tip}}$ is obtained from the average of the gradients from separate linear fits to the points obtained at the three different locations.}
\end{figure}

We then track the tip-induced energy shift by extracting the energy $E_0^{*}$ of the lowest-lying peak in spectra for each tip height. Figures S2(b) \& (c) show examples of spectra acquired at $z_{\mathrm{setpoint}}(\circ) + z_{\mathrm{offset}}$ for each type of surface. The energy of the peak in each spectrum is obtained using fitting of a Lorenztian function with a suitable background. As the peak energy is found to be an approximately linear function of $z_{\mathrm{tip}}(\textbf{\textit{r}})$, the gradients d$E_{0}^{*}$/d$z_{\mathrm{tip}}$ are then obtained using linear fits.

The results of this procedure yield d$E_{0}^{*}$/d$z_{\mathrm{tip}}$ = 0.032~meV/pm for the Type 1 surface [Fig. S2(d)], and d$E_{0}^{*}$/d$z_{\mathrm{tip}}$ = 0.061~meV/pm for the Type 2 surface [Fig. S2(e)]. It is unknown why the TIBB effect has a different strength depending on which surface is being measured.

Approximating the actual sample surface as perfectly flat, we obtain the compensating energy shift
\begin{equation*}
E_{\mathrm{shift}}(\textbf{\textit{r}}) = -\frac{dE_{0}^{*}}{dz_{\mathrm{tip}}} \cdot z_{\mathrm{setpoint}}(\textbf{\textit{r}}).
\end{equation*}
After this energy shift is applied to every pixel, the peaks in the spatially averaged $\mathrm{d}I/\mathrm{d}V$ significantly sharpen. The d\textit{I}/d\textit{V} is then averaged over a 10 meV window around the energy of the lowest-lying peak after the energy shifts. The resulting image represents the spatial distribution of the lowest-lying peak regardless of the energy at which it appears in the raw data.

\subsection{Compensation for tip-induced energy shifts in spectroscopic imaging data.}
After enacting the procedure described above, the conductance maps at each respective surface, shown in Figs. S3(a) and (b) below, is seen to be nearly uniform within each cluster for the reason described in section A. However, it can change abruptly from one cluster to the next. This attests to the strong localization of the orbitals forming the UHB. There is a noticeably greater anisotropy in the pattern for the Type 2 surface than for the Type 1 surface. This may be related to the lower symmetry in the stacking coordination of the uppermost layer of clusters for the Type 2 surface \cite{Butler2020}.

\begin{figure}
\centering
\includegraphics[scale=1]{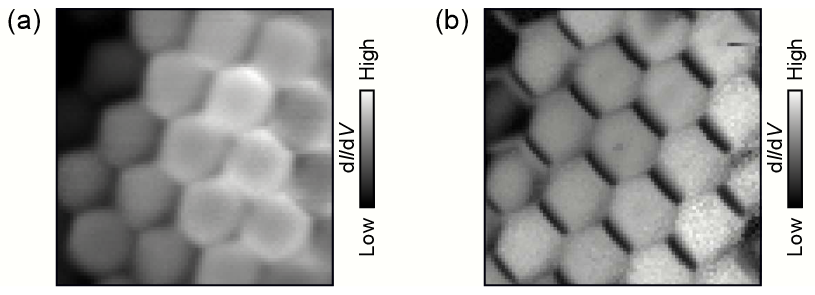}
\caption{\label{fig:3} Spatial distribution of conductance peaks after compensation for tip-induced effects. (a) and (b) The adjusted $\mathrm{d}I/\mathrm{d}V$ data derived from the same spectroscopic imaging data sets shown for the Type 1 \& 2 surfaces, respectively, in Fig. 2 of the main work and Fig. S1 above.}
\end{figure}

\section{Comparison of peaks in conductance spectra acquired at Type 1 \& 2 surfaces.}
In the main text, a clear Franck-Condon progression of peaks due to coupling of a transient charged state to the CDW amplitude mode has been established for the Type 2 surface (unpaired SD clusters). As seen in Fig. 1(d) of the main work, a sharp peak in conductance also appears at the onset of the UHB for the Type 1 surface. The same fitting algorithm described in the main text was applied to data for the Type 1 surface, and the result is shown in Fig. S4(a), alongside that for the Type 2 surface for comparison [Fig. S4(b)]. Here however, the algorithm does not find an equally spaced series of \textit{e-ph} replica peaks. 

What is the difference between the two types of surface? Is the difference qualitative, or only an outcome of tuning some of the parameters governing the lineshape? In Eqn. 2 given in the main text, two parameters can control the formation or suppression of multiple \textit{e-ph} peaks: 1) If the \textit{e-ph} coupling $D$ is small, only the lowest-lying peak manifests. 2) If the peak broadening $\Gamma_{n}$ becomes large for $n > 0$, existing \textit{e-ph} replica peaks may be smeared out such that they are not recognizable. It is unclear from the fitting result shown in Fig. S4(a) whether this latter explanation can be supported.

\begin{figure}[!ht]
\centering
\includegraphics[scale=1]{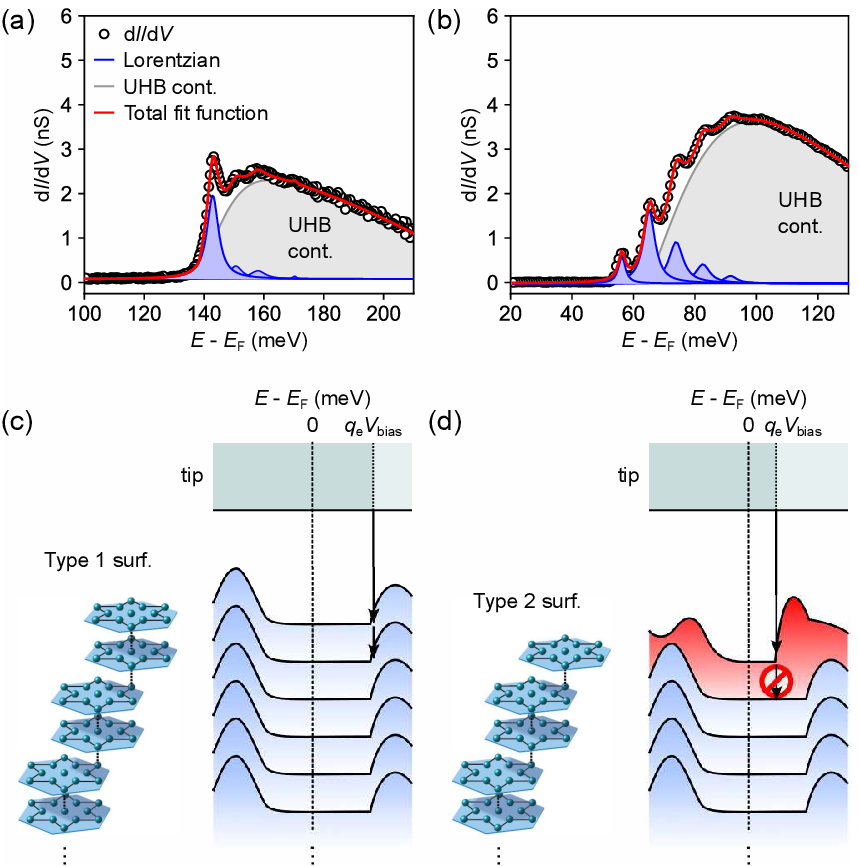}
\caption{\label{fig:4} Comparison of spectra acquired at the Type 1 \& 2 surfaces. (a) Results of fitting to an example curve acquired at a Type 1 surface, using the same algorithm as described in the main text. (b) For comparison, the same fitting to the spectrum acquired at the Type 2 surface presented in the main text. (c) and (d) Qualitative depictions showing electron tunneling into the UHB for the type 1 and 2 surfaces. Assuming that the underlying dimerized bulk electronic structure has an energy gap of $\sim$150~meV, similar to the dimerized Type 1 surface, the lower UHB onset for the Type 2 surface leads to the UHB being decoupled from the bulk, in turn leading to long dwell times on the surface SD clusters.}
\end{figure}

A hint of the mechanism behind the different behaviors is found in the broadening of the lowest-lying peak, $\Gamma_0^{*}$ for each case. For the Type 1 spectrum $\Gamma_0^{*}$ = 4.64~meV, which corresponds to a lifetime of the transient charged state of $\tau _0$ = 71~fs, significantly shorter than the lifetime at the Type 2 surface (152~fs). It is possible that the reduced lifetime of the transient charged state in turn reduces the effective \textit{e-ph} coupling strength, such that no clear Franck-Condon progression develops. This corresponds to an effective reduction of $D$ for the Type 1 surface, suppressing the series of replicas.

The reduced lifetime of $\sim$71~fs for the Type 1 surface is closer to the value estimated for the doublon-hole recombination time in a recent report  by Ligges \textit{et al.} on time-resolved photoemission spectroscopy measurements \cite{Ligges2018}. (We infer that these measurements were acquired on the Type 1 surface based on the observed energy of the UHB at $E - E_{\mathrm{F}} \approx$ 175 meV, and also on the \textit{a priori} likelihood for any given cleaved surface to be of Type 1 \cite{Butler2020}.) The lifetime estimated in the current work may be longer than the lifetime for the excited state of $\sim$20~fs reported by Ligges \textit{et al.} due to a selection bias inherent to our measurement strategy: We make an effort to avoid, as much as possible, impurities and other defects at the surface, and select for SD clusters in the cleanest areas, whereas photoemission spectroscopy measurements necessarily average over a large surface area including many point defects. This may explain the relatively shorter observed doublon-hole recombination time in the photoemission measurements.

The difference in lifetime of the transient charged states at the Type 1 and Type 2 surfaces can be understood by considering the stacking environment and the electronic structure of the both uppermost and underlying TaS$_2$ layers in each case. As proposed recently, the large spectral gap found at the dimerized Type 1 surface likely also corresponds to the bulk spectral gap \cite{Butler2020}, because it is a bulk-like termination of the dimerized three-dimensional CDW pattern. This is depicted in Figs. S4(c) and (d). The lower onset of the UHB at the Type 2 surface, is peculiar to that surface as it likely results from a breaking of the bulk pattern of molecular orbital dimerization. The underlying bulk has a larger gap than the Type 2 surface, and an onset of the UHB at around 140 meV. The UHB at the Type 2 surface, starting at around 60~meV is then electronically decoupled from the underlying layer. An electron dwelling in the bound state of an SD cluster at the Type 2 surface is unable to hop to the underlying layer because that layer has no states at that energy. In contrast, an electron dwelling in a cluster at the Type 1 surface can hop to the UHB at the same energy in the underlying layer. Therefore the transient charged states of the Type 2 surface have enhanced lifetimes, stronger \textit{e-ph} coupling $D$, and a more fully developed series of \textit{e-ph} replicas.

\section{Reproducibility of observation of conductance peaks.}
From the spectroscopic imaging data presented in Fig. S1 above, it is clear that the spectroscopic characteristics of each individual cluster at a given surface are broadly similar. The features reported in this work were robustly reproducible and seen to be ubiquitous over the surface of several cleaved 1\textit{T}-TaS$_{2}$ samples, although it was found that at or near point defects, either the peaks ceased to exist or their appearance became irregular (even including strong enhancement of the peak intensity for some spectra acquired at the Type 1 surface). The data shown in Figs. 1 \& 2 of the main work represent the typical behaviors observed at least a few nanometers away from defects or CDW domain walls. 

A single example of a typical spectrum acquired on a cluster at the Type 2 surface is analyzed in the main text. In Figs. S5 (a)--(d), analyses are presented for conductance spectroscopy measurements acquired at four different locations on the same sample with the Type 2 surface termination. Aside from small variations, the behavior is seen to be similar. The parameters $E_{\mathrm{ph}}$, $D$ and $\Gamma_{0}^{*}$, extracted using the same fitting algorithm as described in the main text, are presented in Table I.

\begin{figure}[!ht]
\centering
\includegraphics[scale=1]{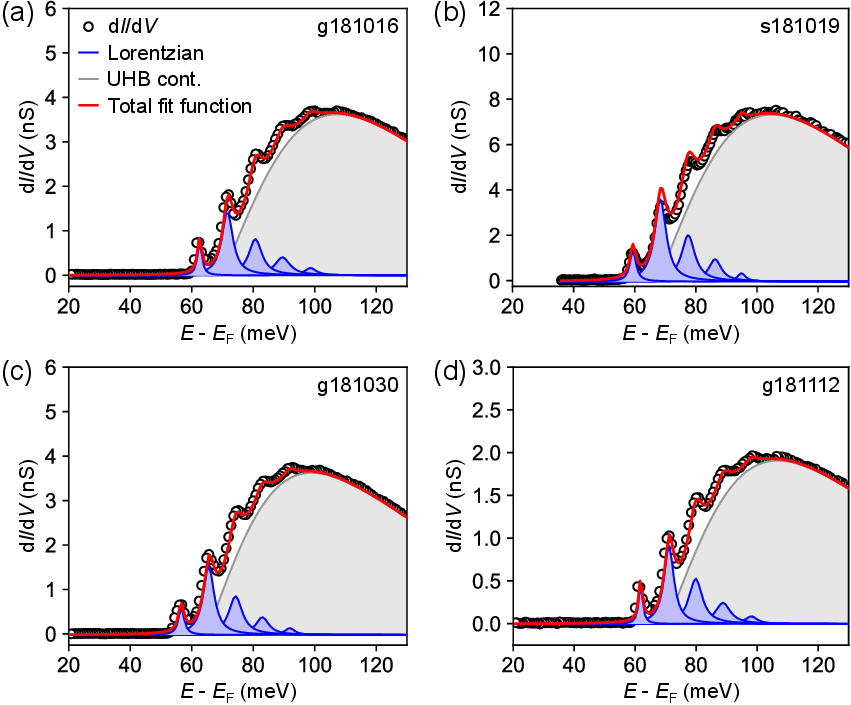}
\caption{\label{fig:5} (a)--(d) Additional analyses of \textit{e-ph} replicas at different sample locations. The valued for $E_{\mathrm{ph}}$, $D$, and $\Gamma_{0}^{*}$ are given in Table I.}
\end{figure}

From the additional data presented in Table 1, it can be seen that there is a weak negative correlation between the fitted values of $D$ and $\Gamma_0^{*}$, i.e. there is a positive correlation between $D$ and $\tau_0$. This is consistent with the idea that a longer dwell time for the injected electron on the cluster results in stronger electron-phonon coupling.

\begin{table}[!h]
  \begin{center}
    \caption{Fitted values for spectra in Fig. S5.}
    \label{tab:table1}
    \begin{tabular}{c c c c}
      \hline
      \hline
      Data set & \quad $E_{\mathrm{ph}}$ (meV) & \quad $D$ & \quad $\Gamma_{0}^{*}$ (meV) \\
      \hline
      g181016 & \; 9.00 & \; 1.80 & \; 1.67 \\
      s181019 & \; 8.89 & \; 1.73 & \; 2.38 \\
      g181030 & \; 8.80 & \; 1.75 & \; 2.17 \\
      g181112 & \; 9.03 & \; 1.84 & \; 1.52 \\
      \hline
    \end{tabular}
  \end{center}
\end{table}

\newpage

\section{Spectroscopic imaging near defects.}

For the interpretation of the following observations, our premise is that the narrow peaks arise from interactions between an injected electron (doublon) which dwells in a cluster for a finite time ($\sim$100 fs), and, at least for the Type 2 surface, excites amplitude mode phonons. If for any reason the dwell time of the electron becomes too short, or if a defect of the ionic lattice results in the amplitude mode not being locally supported (most obviously, at a domain wall in the CDW), the peaks should be significantly broadened, or disappear entirely. This expectation is satisfied by the observations below.

\begin{figure}
\centering
\includegraphics[scale=1]{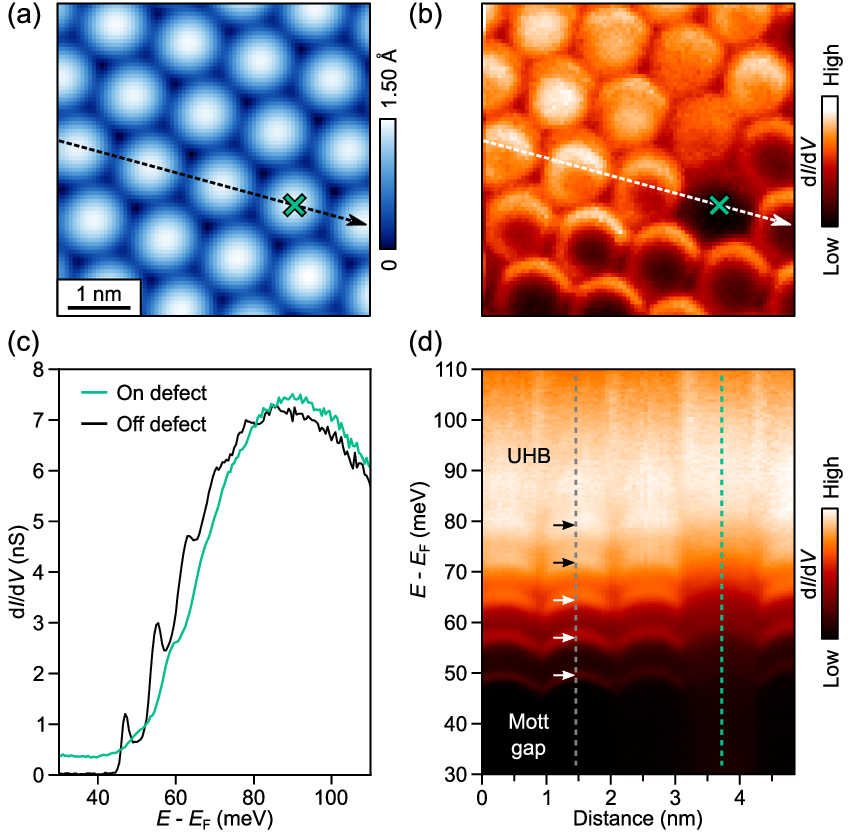}
\caption{\label{fig:6} Spectroscopic imaging around a defect at the Type 2 surface. (a) A constant-current topograph (setpoint $V$ = 0.11~V, $I$ = 0.44~nA). (b) A selected conductance map, at 65~mV, showing the darkening of a single SD cluster induced by a point defect [marked with a green cross in both (a) and (b)]. (c) Single-point $\mathrm{d}I/\mathrm{d}V(V)$ spectra on (black curve) and away from the defect (green curve). (d) A spectroscopic linecut sampled along the dashed lines in (a) and (b). The black and green vertical lines correspond to the point spectra shown in (c). Arrows mark the five peaks under discussion.}
\end{figure}

Figure S6 shows a point defect which impacts one particular SD cluster. This has the effect of drastically smearing out the peaks in that cluster [as seen in panels (c) and (d)], as well as creating some residual density of states below the onset energy of the UHB. We see that the suppression or smearing of the peaks is confined to a single cluster, meaning the conditions giving rise to the peaks are determined locally.

\begin{figure}
\centering
\includegraphics[scale=1]{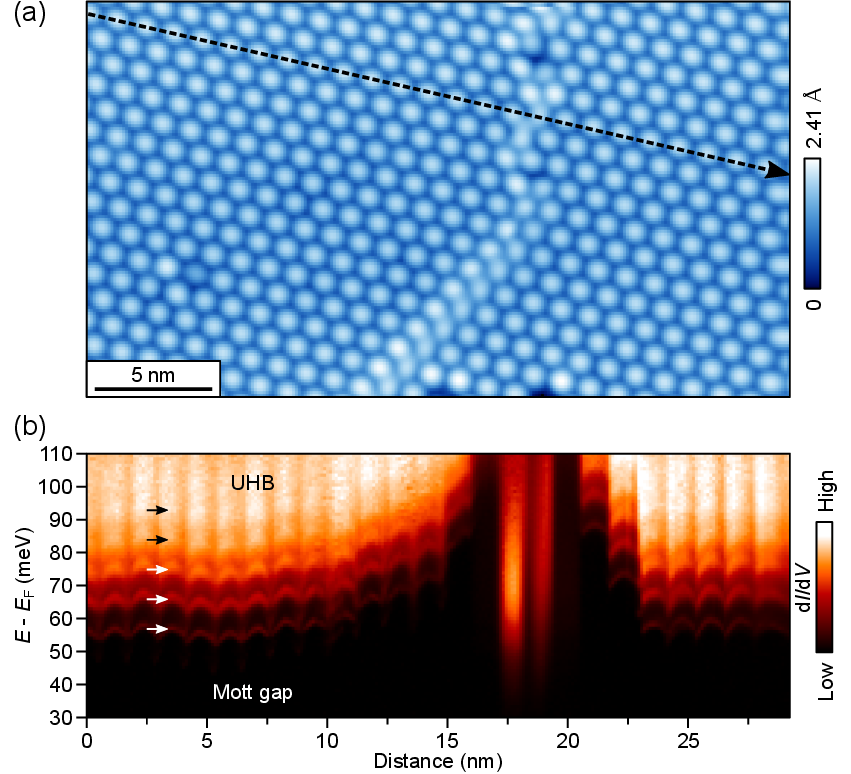}
\caption{\label{fig:7} Spectroscopic imaging near a CDW domain wall at the Type 2 surface. (a) A constant-current topograph ($V$ = 0.25~V, $I$ = 1~nA) showing a domain wall extending from the top to the bottom of the field of view. (b) A spectroscopic linecut sampled on a path marked by the black dashed line in (a). Black and white arrows mark the peaks under discussion.}
\end{figure}

Figure S7 shows the effect of a CDW domain wall. The series of peaks, shown in the spectroscopic linecut of panel (b), persists until about 2~nm away from the domain wall. Although they suffer from energy shifts associated with the Coulomb potential landscape, we see that these shifts are rigid and the peak spacing is approximately maintained (i.e., $E_{\mathrm{ph}}$ remains the same). The peaks disappear at the domain wall, where the Mott gap partially collapses, as is well-known \cite{Cho2017}, and the amplitude mode phonon is not expected to be supported by the local lattice structure.

\begin{figure}
\centering
\includegraphics[scale=1]{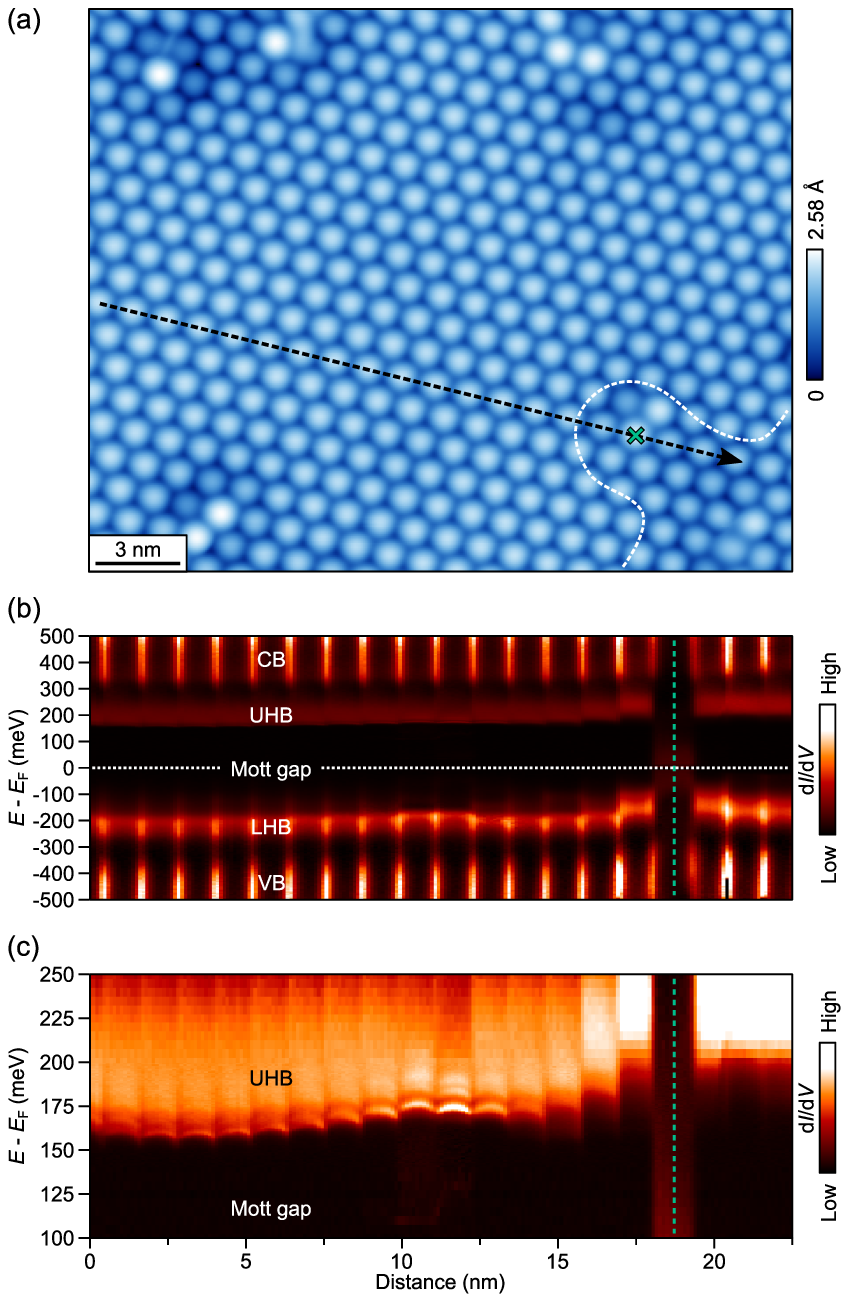}
\caption{\label{fig:8} Spectroscopic imaging on approach to a cluster of defects at the Type 1 surface. (a) A topograph ($V$ = 0.25~V, $I$ = 1~nA) showing several defects surrounding a pristine region. The white dashed line encloses a small region where the defects influence the electronic structure, and the green cross marks a defect at which the Mott gap is seen to collapse in the spectroscopic linecut shown in (b), sampled along the path marked by the black dashed line in (a). The location of the same defect is indicated with a green dashed line. (c) A zoom-in view on the same linecut shown in (b), showing the spectroscopic peak at the UHB edge on approach toward the defect-rich area.}
\end{figure}

A similar observation at the Type 1 surface is shown in Figure S8. Here we see that the sharp peak exists everywhere until a few nanometers away from a prominent local defect, at which the Mott gap collapses. As the topograph shows that SD reconstruction is well-formed everywhere, even at the defects, the breakdown of the peak near the defect is likely purely electronic in nature. For example, the lifetime of the excitation could be drastically shortened by impurity-mediated hopping.

\section{Further discussion of the observed phonon energy.}

Tentatively attributing our observed phonon energy to the 2.4~THz amplitude mode excitation, we note that the observed peak spacing $E_{\mathrm{ph}} \approx$ 9~meV is 9$\sim$10\% smaller than the previously reported energy of the amplitude mode. We suggest a number of possible reasons for this discrepancy. The first is that a renormalized phonon energy for highly excited states, as compared with the lowest excited or ground state, is possible. The second possibility relates to observations by Stojchevska \textit{et al.} \cite{Stojchevska2014} that the amplitude mode frequency appears to be reduced by a few percent upon laser pulse-induced switching into a ``hidden'' metastable state of the CDW. However it is unclear whether the hidden state has any relation to the Type 2 surface CDW stacking configuration observed here.
The third is that the highly localized phonon excitation observed here (localized around a point in the plane of the surface, and also to the uppermost layer), may differ in energy from the case of a photo-excitation induced amplitude mode occurring coherently over a wide area, and throughout the multiple layers probed by photo-excitation measurements.
Finally, while we observe this phononic or vibronic excitation in a SD cluster at the Type 2 surface, it is likely that previously reported measurements either were conducted only on the dimerized surface we identify as Type 1 \cite{Butler2020}, or lack the surface specificity to resolve Type 2 behavior even if conducted at a Type 2 surface. It is possible that the energy of the amplitude mode may subtly depend on the stacking configuration of the host SD clusters, and this may explain the small energy discrepancy.